\documentclass[10pt]{article}

\usepackage{epsfig,amsfonts,amssymb,newlfont,rotate,float,array}

\begin{document}

\title{\bf A measure of conductivity 
for lattice fermions\\ at finite density.}
\author{J.~L.~Alonso$^1$\footnote{Electronic address: 
\tt buj@gteorico.unizar.es}\,,
L.~A.~Fern\'andez$^2$\footnote{Electronic address: 
\tt laf@lattice.fis.ucm.es}\,, and
V.~Mart\'{\i}n-Mayor$^{3}$\footnote{Electronic address: 
\tt Victor.Martin@roma1.infn.it}\,.\\
\\
\normalsize$^1$ \it Departamento de F\'{\i}sica Te\'orica, Universidad de
Zaragoza. 50009 Spain.\\
\normalsize$^2$ \it Departamento de F\'{\i}sica Te\'orica I, 
Universidad Complutense de Madrid. 28040 Spain.\\
\normalsize$^3$ \it Dipartimento di Fisica, 
Universit\'a di Roma I, 00185 Italy.
}
\date{\today}
\maketitle
\begin{abstract}
\small We study the linear response to an external electric field of a
system of fermions in a lattice at zero temperature. This allows to
measure numerically the Euclidean conductivity which turns out to be
compatible with an analytical calculation for free fermions. The
numerical method is generalizable to systems with dynamical
interactions where no analytical approach is possible.

\medskip


\end{abstract}

\section{\protect\label{S_INT}Introduction}

The study of transport properties in a system of charged fermions is
an interesting subject in areas as different as Physics of Plasma,
Quantum Chromodynamics or Metals. In particular, the measure of the
electrical conductivity is a very difficult and yet interesting
problem, specially in presence of nonperturbative effects. In such a
case, numerical methods are called for.  In this work, we want to show
that lattice-regularized Euclidean field theories can be useful in
this respect, at least in the limit of vanishing temperature. However,
it should be emphasized that the so called {\em sign-problem} needs
still to be overcome in many cases~\cite{LOMBARDO} (see, though,
Refs.~\cite{FOURFERMIONS,SU2} for some successful simulations at finite
density).

In this paper, we restrict ourselves to a model consisting of fermions
that only interact with an external electromagnetic field. In spite of
its simplicity, it shares many properties with more realistic models,
and it can be considered as a necessary first step to check any
numerical method which could be used in models with dynamical
interactions. We consider the standard $U(1)$ lattice-action, with
Wilson fermions and finite chemical potential, but with the gauge
variables held fixed.  We shall study the residue of
the pole of the electrical conductivity at zero-frequency, which is
purely imaginary.  Since a non-vanishing value for this residue
unambiguously signals a conducting phase, this is a rather interesting
quantity in our opinion. In order to obtain it, we measure the
electrical-current induced in the system by an external electric
field. This technique requires a numerical calculation even in the
case of an external spatially-homogeneous time-dependent
electromagnetic field. The delicate point, however, is that our
electric field varies in Euclidean time. One can nevertheless
assume that there is a linear relation between the Euclidean current
and the Euclidean electric field, at least for small fields. This {\em
Euclidean conductivity} presents a pole whose residue can be
straightforwardly measured. To check that the obtained result is
physical, we follow a very elegant procedure due to
Kohn\cite{KOHN}. He showed that the {\em real-time} residue can be
measured by studying the sensitivity of the ground-state energy to an
external Aharonov-Bohm electromagnetic field. We show how can this be
done in the lattice formalism, and, in this particularly simple case
of free fermions, we calculate it (unfortunately, the Kohn recipe
seems really hard to use in a Monte Carlo study of a self-interacting
problem).  Although at present we lack a rigorous proof of the
equivalence of both calculations, its excellent numerical agreement
gives a strong support to the linear response method.

\section{\protect\label{MODEL}The Model}

Let us consider a model of Wilson Fermions\cite{WILSON,WILSON2} 
in a lattice of spacing $a_\mathrm{s}$ in the three spatial directions and
$a_\mathrm{t}$ in the temporal one, coupled to an external 
electromagnetic field. We denote by $\lambda$ to the quotient
$a_\mathrm{t}/a_\mathrm{s}$.
The partition function can be written as
(the $*$ superscript stands for complex 
conjugation)~\cite{POTQUIMH,POTQUIMK,BENDER}
\begin{equation}
{\cal Z}[U]=\int\prod_z\,{\mathrm d}\Psi_z {\mathrm d}\overline{\Psi}_z
\exp\left[\sum_{x,y}\overline{\Psi}_x 
M_{xy}(U)\Psi_{y}\right]\label{PATHINT}\,,
\end{equation}

\begin{eqnarray}
M_{xy}(U)
&=&                
{\mathrm e}^{\lambda\mu} U_{x,0}(\gamma_0-r_\mathrm{t})\delta_{y,x+\hat 0}-
{\mathrm e}^{-\lambda\mu} U_{x,0}^*(\gamma_0+r_\mathrm{t})\delta_{y+\hat 0,x}
\label{MATRIZ}\\
&+&\lambda\sum_{i=1}^3
[U_{x,i}(\gamma_i-r_\mathrm{s})\delta_{y,x+\hat \imath}
-U_{x,i}^*(\gamma_i+r_\mathrm{s})\delta_{y+\hat \imath,x}]+
[(2m+6r_\mathrm{s})\lambda+2r_\mathrm{t}]\delta_{x,y}\ ,\nonumber
\end{eqnarray}
where $U_{x,\nu}={\mathrm e}^{{\mathrm{i} A_{x,\nu}}}$, $A$ being the
gauge field, and $\Psi_x$, $\overline\Psi_x$ are the anticonmuting
Grassmann fermionic fields. The indices $x,y$ run on the points of the
four-dimensional space-time lattice.  We impose periodic boundary
conditions for the gauge field, and periodic in space but antiperiodic
in time ($\nu=0$) for the Grassmann field. The site $x+\hat\nu$ is the
neighbor of $x$ in the $\nu=0,1,2,3$ direction.  For finite temporal
length, $L_0$, the system is at finite temperature $T=(a_\mathrm{t}
L_0)^{-1}$. In this paper we will only consider the zero temperature
($L_0\to\infty$) limit. We follow the prescription of introducing the
chemical potential through an imaginary gauge field $A=(-{\mathrm
i}\lambda\mu,0,0,0)$~\cite{POTQUIMH,POTQUIMK}, which is fairly
convenient for analytical calculations.  

The Wilson parameter, $r$, can be taken different for the
spatial and time directions.
In the limit $\lambda\to 0$ with $a_\mathrm{s}$ fixed the model
describes a spatial lattice with a continuous time (as
electrons in a metal), while for a continuum field theory both spatial
and time continuum limits should be taken.

We shall use the following representation for the (Euclidean)
gamma matrices
\begin{equation}
\gamma_0=\left(\begin{array}{cc}0&1\\1&0\end{array}\right),\quad
\gamma_i=\left(\begin{array}{cc}0&-{\mathrm i}
\sigma_i\\{\mathrm i}\sigma_i&0\end{array}
\right),\
\end{equation}
where $\sigma_i$ are the Pauli matrices.

To define the electric four-current in the lattice we recall that
in the space-time continuum limit it is defined as
\begin{equation}
j_\nu(x)=\overline{\Psi}(x)\gamma_\nu\Psi(x)\,,
\end{equation}
that can be obtained as a logarithmic derivative of the partition
function respect to the gauge-field.
This calculation can be exactly mimicked on the lattice noticing
that a change in the link variable should be of the form
$U_{x,\nu}\rightarrow {\mathrm e}^{{\mathrm i} \alpha_{x,\nu}}U_{x,
\nu}$.  In this way one obtains~\cite{POTQUIMK}:
\begin{equation}
\langle j_{x,\nu}\rangle={\mathrm i}\,\frac{\partial \log {\cal Z}}
{\partial\alpha_{x,\nu}}\ ,
\end{equation}
where now
\begin{eqnarray}
\langle j_{x,0}\rangle&=&\left\langle \overline{\Psi}_x {\mathrm
e}^{\lambda\mu} U_{x,0} (\gamma_0-r_\mathrm{t})\Psi_{x+\hat 0} +
\overline{\Psi}_{x+\hat 0} {\mathrm e}^{-\lambda\mu}
U_{x,0}^*(\gamma_0+r_\mathrm{t}) \Psi_{x}\right\rangle\,\label{J0}\\
\langle j_{x,i}\rangle&=&\lambda\left\langle \overline{\Psi}_{x}
U_{x,i}(\gamma_i-r_\mathrm{s})\Psi_{x+\hat \imath}+
\overline{\Psi}_{x+\hat \imath} U_{x,i}^*(\gamma_i
+r_\mathrm{s})\Psi_{x}\right\rangle \,,\ i=1,2,3\, .\label{JI}
\end{eqnarray}
The $j_0$ component is just the electric charge density that one
encounters by differentiating with respect to $\lambda\mu$ the free 
energy density~\cite{POTQUIMK}. Moreover, from the gauge
invariance of the determinant of the fermionic matrix, $M$,
it is straightforward to prove the lattice continuity equation, for
any configuration of the electromagnetic-field:
\begin{equation}
0=\sum_\nu \left(\langle j_{x,\nu} \rangle - \langle j_{x-\hat\nu,\nu}
\rangle\right)\, .
\end{equation}

Eqs.~(\ref{J0}) and (\ref{JI}) can be written free of Grassmann
variables as
\begin{eqnarray}
-\langle j_{x,0}\rangle&=&{\mathrm e}^{\lambda\mu} U_{x,0} \mathrm{Tr} 
[ (\gamma_0-r_\mathrm{t})
 M^{-1}_{x+\hat 0,x}]+{\mathrm e}^{-\lambda\mu} U^*_{x,0} \mathrm{Tr} 
[ (\gamma_0+r_\mathrm{t}) M^{-1}_{x,x+\hat 0}]\ , \label{J0NG}\\
-\langle j_{x,i}\rangle&=&\lambda U_{x,i} \mathrm{Tr} 
[ (\gamma_i-r_\mathrm{s}) M^{-1}_{x+\hat \imath,x}]
+\lambda U^*_{x,i} \mathrm{Tr} [ (\gamma_i+r_\mathrm{s})
 M^{-1}_{x,x+\hat \imath}]\ ,
\end{eqnarray}
where Tr stands for the trace over Dirac indices.
The above expressions and the relation 
\begin{equation}
M(U^*) = \gamma_1\gamma_3 \left(M (U)\right)^* \gamma_3\gamma_1\,,
\end{equation}
allow to prove that
\begin{equation}
\langle j_{x,\nu} \rangle^*_U = \langle j_{x,\nu} \rangle_{U^*}\ .
\label{JCONJ}
\end{equation}
In an uniform electrical field, the charge density should remain
constant under field inversion, while the electrical current should
change sign. Therefore, from Eq.~(\ref{JCONJ}) one expects the former
to be real and the latter to be imaginary (Euclidean space-time!).

\begin{figure}[t!]
\begin{center}
\epsfig{file=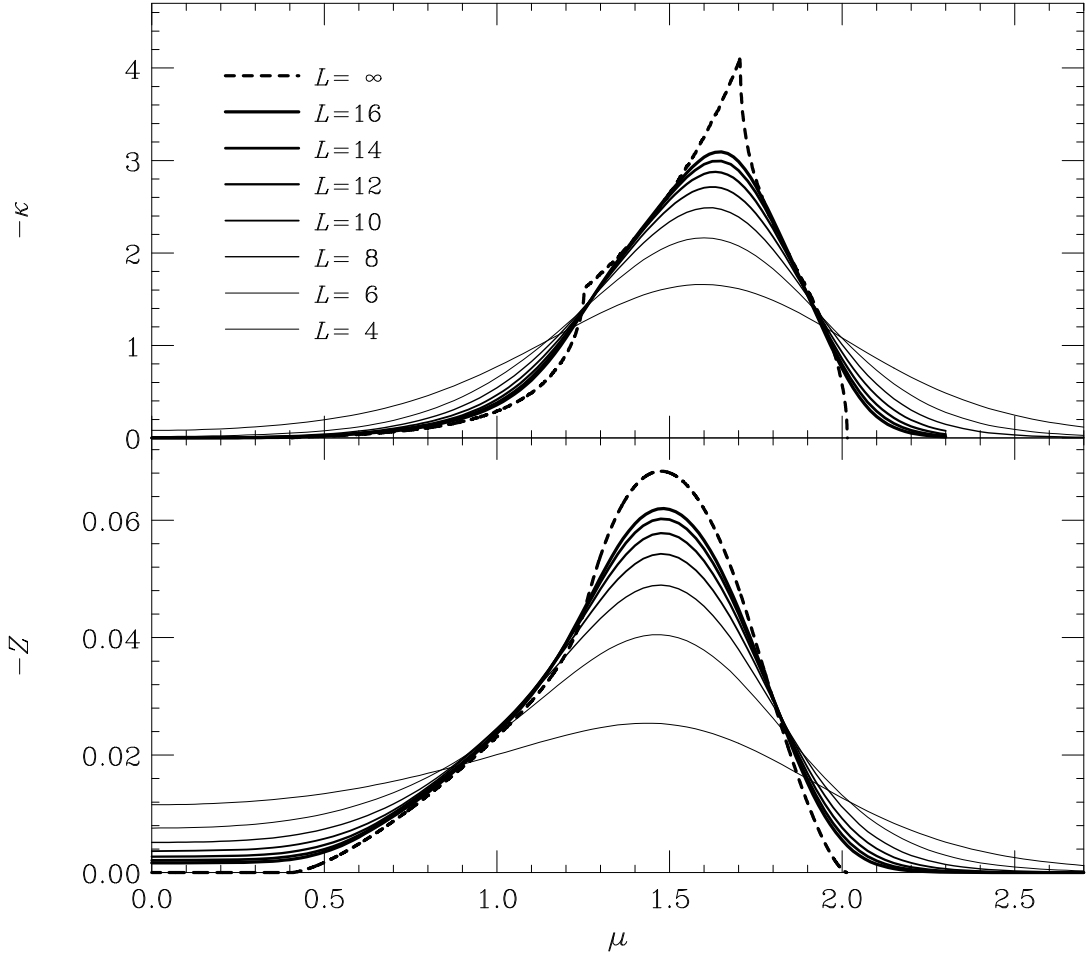,width=0.8\linewidth,angle=90}
\end{center}
\caption{\small Density of states (upper part) and residue of the
conductivity (lower part) for $\mu>0$. 
The dashed lines correspond to analytical
calculations in Fourier space (see Eqs. (\ref{RHO}), (\ref{KAPPA}),
and (\ref{CONDKOHN})). The solid lines are obtained numerically in
finite lattices (see section \ref{NUMCAL}).  In all cases $r=1$.}
\label{DEN-RES}
\end{figure}

In absence of external fields ($U=1$) the matrix $M$ can be diagonalized 
in Fourier space, which allows to explicitly perform the functional 
integrals, and to compute the free energy and the propagator.  
For brevity, we only quote the result for the 
charge density in the case $r_\mathrm{t}=r_\mathrm{s}=1$, $\mu>0$, that in
the infinite volume limit reads (see 
Refs.~\cite{MONTVAYMUNSTER,GAVAI} for similar calculations),
\begin{equation}
\rho(\lambda,\mu) = 2 \int_{-\pi}^\pi \frac{{\mathrm
d}^3\mbox{\boldmath $k$}}{(2\pi)^3}\, \theta\left(\mu -
\lambda^{-1}E(\mbox{\boldmath $k$})\right)\, ,\label{RHO}
\end{equation}
where
\begin{eqnarray}
E(\mbox{\boldmath $k$}) &=& \mathrm{arccosh}\left[
\frac{1+\lambda^2\sum_j \sin^2k_j
+\left(\lambda\Sigma(\mbox{\boldmath $k$}) +1\right)^2}
{2\left(\lambda\Sigma(\mbox{\boldmath $k$}) +1\right)}\right]\,,\\
\Sigma(\mbox{\boldmath $k$})&=& m+\sum_j\,(1-\cos k_j)\ .\label{SIGMA}
\label{ER1}
\end{eqnarray}
An useful quantity is the mechanical compressibility that, at zero-temperature
coincides with the density of states:
\begin{equation}
\kappa(\lambda,\mu)=\frac{\partial \rho(\lambda,\mu)}{\partial \mu}=
 2\int_{-\pi}^\pi \frac{{\mathrm
d}^3\mbox{\boldmath $k$}}{(2\pi)^3}\, \delta\left(\mu -
\lambda^{-1}E(\mbox{\boldmath $k$})\right)=
2\lambda\int_{E(\mbox{\scriptsize\boldmath $k$})=\lambda\mu}
\frac{{\mathrm d}^2 S}{(2\pi)^3}\, \frac{1}{\left\Vert\mbox{\boldmath
$\nabla_k$} E(\mbox{\boldmath $k$})\right\Vert}\,.\label{KAPPA}
\end{equation}
The density of states of the system present a typical band structure
(see the upper part of Fig.~\ref{DEN-RES}, dashed line). The upper limit
of the band corresponds to the saturation due to the Fermi statistics
(one particle per lattice-site). Since the function $E(\mbox{\boldmath
$k$})$ is periodic, its gradient has zeroes in the Brillouin zone, 
producing  non-analiticies as the cusps in Fig.~\ref{DEN-RES}
(the so-called Van-Hove singularities \cite{ASCROFHTMERMIN}).

\section{\protect\label{ELECT}The Electrical Conductivity}

In a classical paper, Kohn~\cite{KOHN} developed an elegant
characterization of a conductor, at zero temperature.
His method allows the measurement of the following limit
for the imaginary part of the electrical conductivity, 
$\sigma''$,
\begin{equation}
Z=\lim_{\omega\to 0}
\omega\sigma''(\omega)\ .
\end{equation} 
If this limit turns out to be non zero, the system is  a conductor.
The construction is as follows. The system of interest is
constrained to verify periodic boundary conditions in the (say)
first spatial direction, and immersed in an Aharonov-Bohm like electromagnetic
field $A=(0,\alpha,0,0)$. 
With this choice of boundary conditions
the product $L_1 a_\mathrm{s} \alpha$ 
is gauge invariant since it represents the
magnetic flux traversing the system. 
It can be shown that
\begin{equation}
Z= -\frac{1}{V_{\mathrm{s}}}\left.\frac{\mathrm d^2 
E_0}{\mathrm d
\alpha^2}\right|_{\alpha=0}\, ,
\end{equation}
where  $E_0$ is the
ground-state energy and $V_{\mathrm{s}}$ is the spatial volume. It is
crucial that the infinite limit volume is taken {\it after} the
$\alpha$ derivative is performed, since the effect of the
Aharonov-Bohm field can be thought of as a change in the boundary
conditions (see below). In the infinite volume limit, the energy no
longer depends on $\alpha$.

In our case, as the free energy and the ground-state energy coincide
in the zero temperature limit,  we can study the residue in
the following way
\begin{equation}
Z=-\lim_{V_\mathrm{s}\to \infty} \lim_{T\to 0}\left.\frac{\mathrm
d^2 \tilde f}{\mathrm d \alpha^2}\right|_{\alpha=0}\,,
\end{equation}
where $\tilde f$ is obtained from the intensive free energy $f$ after
subtracting the vacuum contribution: $\tilde f(\mu)=f(\mu)-f(0)$.

Let us sketch the calculation.
The free energy should be calculated in a finite
volume and at finite temperature.
We introduce our system in the Aharonov-Bohm electromagnetic
field:
\begin{equation}
U_{x,0}=U_{x,2}=U_{x,3}=1\ ,\  U_{x,1}={\mathrm e}^{{\mathrm i}\alpha}\, .
\end{equation}
This field can be transformed into a boundary effect by performing
the following gauge transformation:
\begin{equation}
U_{x,\nu}\rightarrow U^{\mathrm G}_{x,\nu}={\mathrm e}^{{\mathrm i} g(x)}\, U_{x,\nu} 
{\mathrm e}^{-{\mathrm i} g(x+\hat\nu)}\,\ ,\ g(x)=\alpha x_1 \, ,
\label{CAMBIOGAUGE}
\end{equation}
so that $U^{\mathrm G}=1$ excepting
\begin{equation}
U^{\mathrm G}_{(x_1=L_1-1),1}={\mathrm e}^{{\mathrm i} \alpha L_1}\, .
\end{equation}
By direct inspection of the fermion matrix in Eq.~(\ref{MATRIZ}),
one can easily recognize that a system verifying periodic boundary
conditions in the 1 direction in the field $U^{\mathrm G}$ is equivalent
the same system with no field at all, but verifying
\begin{equation}
\Psi(x_0,x_1+L_1,x_2,x_3)={\mathrm e}^{{\mathrm i} \alpha
L_1}\Psi(x_0,x_1,x_2,x_3)\ ,
\end{equation}
This amounts to substituting $k_1$ by $k_1+\alpha$ in the momentum-quantification in a finite lattice. For a system of free fermions the 
free energy can be now straightforwardly calculated.  Once the $\alpha$ derivative
is performed, the zero temperature limit can be taken by transforming
the $k_0$ sum into an integral. We get, 
in the simplest case $r_\mathrm{t}=r_\mathrm{s}=1$, $\mu>0$, 
\begin{equation}
Z=
-2\int_{-\pi}^\pi\frac{\mathrm d^3 \mbox{\boldmath $k$}}{(2\pi)^3}
\frac{ \partial^2E(\mbox{\boldmath $k$})}{\partial k_1^2} 
\theta(\mu - \lambda^{-1}E(\mbox{\boldmath $k$}))\,.
\label{CONDKOHN}
\end{equation}

Notice that for the empty system, $\mu < \lambda^{-1}E_\mathrm{min}$,
the integral vanishes, as well as for the full band $\mu >
\lambda^{-1}E_\mathrm{max}$, since $E(\mbox{\boldmath $k$})$ 
is a periodic function of $k_1$.  The three dimensional integrals 
(\ref{CONDKOHN}) can be
performed using a Monte Carlo method. The results are shown in
Fig.~\ref{DEN-RES} (dashed line in the lower part). 

\section{\protect\label{NUMCAL}Numerical Calculations}

In this section, we are going to reproduce the results of the sections
\ref{MODEL} and \ref{ELECT} by directly considering the integration of
the partition function. This method has the advantage of being
generalizable to inhomogeneous external fields, and also when
interacting dynamical fields are present. Examples of how
to introduce an external field on an interacting lattice-gauge system
can be found in Refs.~\cite{EXTERNAL}. To compute the partition
function it is necessary to work in finite lattices, consequently, 
an infinite volume limit should be taken.

We have carried out measures in symmetric lattices of sizes 
$L=4,6,8,10,12,14$ and $16$, with $m=1/2$ and $\lambda=1$.
For the hopping term, we have taken $r_\mathrm{s}=r_\mathrm{t}=1$. 
As the integral in the fermionic fields is
Gaussian, the computation of the electric current just requires the
inversion of a $4V$ matrix, $V$ being the space-time volume. The 
fermion matrix (\ref{MATRIZ}) being sparse, we
have used a conjugate-gradient algorithm for the numerical inversion.

We first consider the density of states in a vanishing external field.
In order to measure $\partial \rho/\partial \mu$ we invert the matrix
at $\mu\pm\epsilon$ for $\epsilon$ small enough. In interacting
systems the derivative can be calculated in terms of connected
correlation-functions~\cite{MONTVAYMUNSTER}.

The numerical results are plotted in Fig.~\ref{DEN-RES}, upper part, 
together with
the infinite volume values obtained analytically.
Although the finite size effects are non negligible even in
the larger lattices for most values of $\mu$, there is a clear trend
to the asymptotic values.

Unfortunately, for an interacting system it is not immediate how to
implement Kohn's method for calculating the residue of the
conductivity. In fact, the free energy is rather hard to calculate
with a Monte Carlo simulation and what one directly obtains are
mean-values. We are now going to present a different way of computing
the residue, by directly measuring the system response to an external
electrical field.  Notice that the presence of an electric field
requires a non-homogeneous vector potential and consequently the
inversion of the fermion matrix can no longer be performed in closed
analytical form.  This new recipe can be straightforwardly
generalized to interacting systems, but its equivalence with the
Kohn's method  is just an
ansatz. Nevertheless the agreement is excellent, as we will show.

By analogy with continuum
electrodynamics, we want to study the electric current induced in the
system by an external weak uniform electric field in the $1$ direction.
The conductivity (in the frequency domain), will be the proportionality
constant between the electrical current and the external field.

There are some subtleties that need to be considered when putting an
external electric field on the lattice.  We take the gauge-field
configuration ($t=x_0$)
\begin{equation}
U_{x,0}={\mathrm e}^{{\mathrm i}\, {\cal E}_t x_1},\quad U_{x,i}=1\,, 
\label{UELECTRICO}
\end{equation}
\begin{equation}
{\cal E}_t=\frac{2\pi}{L_1} n_t\ ,\quad
n_t\in\left\{ -\frac{L_1}{2},-\frac{L_1}{2}+1,\ldots,
\frac{L_1}{2}-1,\frac{L_1}{2}\right\}\,.
\end{equation}
Notice that the quantization of the electric-field is due to the
spatial boundary conditions.
To preserve the translational symmetry, the displaced gauge field
\begin{equation}
U_{x,0}={\mathrm e}^{{\mathrm i}\, {\cal E}_t (x_1-\xi)}\ ,\quad \xi\
\mathrm{integer}\ ,
\end{equation}
should be a gauge-transform of the one in Eq.~(\ref{UELECTRICO}).
Since the needed gauge transformation is analogous to
Eq.~(\ref{CAMBIOGAUGE}), it is easy to check that the condition that
allows this transformation is the trivialness of the Polyakov loop:
\begin{equation}
\prod_{t=0}^{L_0-1}U_{(t,\mbox{\scriptsize\boldmath $x$}),0}=1\quad
\mathrm{or}\quad \sum_{t=0}^{L_0-1} {\cal E}_t = 2\pi n\, ,
\label{POLYAKOV}
\end{equation}
with $n$ integer.  This condition also allows to transform the gauge
field to the Coulomb gauge $A_0=0$.  If condition (\ref{POLYAKOV}) is
violated, the translational invariance is lost and the electric
current is no longer spatially homogeneous even on a homogeneous
electric-field. However, with the correct field choice
(\ref{POLYAKOV}), we get a homogeneous electrical current aligned with
the external electrical field, and imaginary as anticipated in
Eq.~(\ref{JCONJ}). In order to directly compare with the results
obtained with Eq.~(\ref{CONDKOHN}), let us define
\begin{equation}
j(t)={\mathrm i}\langle j_{x,1}\rangle
\end{equation}
If we want to stay within linear-response theory, we have
to postulate a linear relation between the Fourier transform of the
electrical current $j(t)$ and the external electrical field ${\cal E}_t$:
\begin{equation}
\hat \jmath(\omega)=\sigma(\omega)\hat {\cal E}(\omega)\, .
\end{equation}
Notice that both $j(t)$ and ${\cal E}_t$ being real, 
$\sigma(-\omega)=\sigma^*(\omega)$.
However, the results can be more cleanly cast in terms of a modified
Fourier transform for the electrical field:
\begin{equation}
{\tilde {\cal E}}(\omega)= \frac{1}{\sqrt{L_0}}\sum_{t=0}^{L_0-1}
{\cal E}_t{\mathrm e}^{-{\mathrm i}\omega(t+1/2)}\, .
\end{equation}
The rationale for this is that the electrical field ${\cal E}_t$ on the
lattice lives mid-way between sites at times $t$ and $t+1$. The
modified conductivity $\tilde \sigma(\omega)= \hat
\jmath(\omega)/{\tilde {\cal E}}(\omega)$  is related with the previous one
by
\begin{equation}
\tilde \sigma(\omega)=\sigma(\omega) {\mathrm e}^{{\mathrm
i}\omega/2}\, .
\end{equation}
The nice feature of $\tilde\sigma(\omega)$ is that it turns out to be
purely imaginary. 

\begin{figure}[t!]
\begin{center}
\leavevmode
\epsfig{file=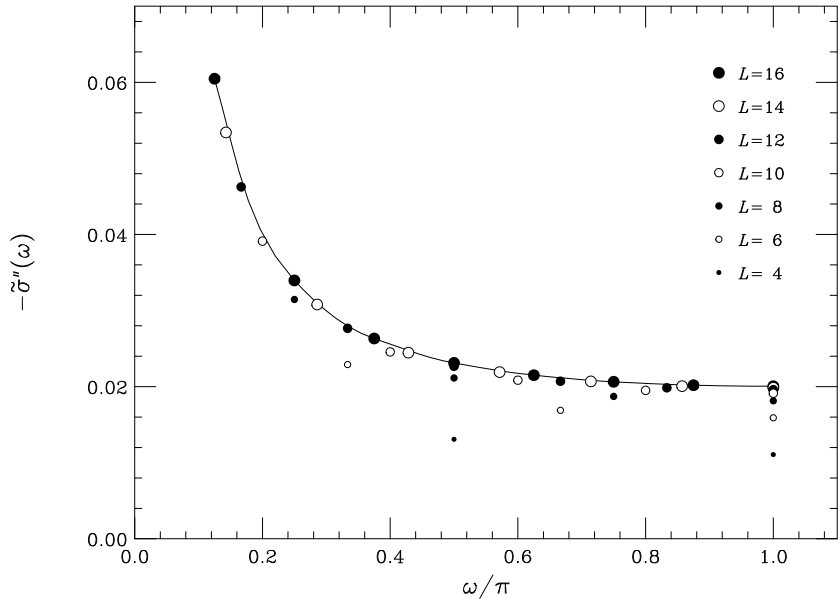,width=0.5\linewidth,angle=90}
\end{center}
\caption{\small The imaginary part of the conductivity $\tilde\sigma(\omega)$
of a system of free Wilson fermions at $r=1$ , $m=1/2$ and $\mu=1$.}
\label{CURVACONDUCTIVIDAD}
\end{figure}

In Fig.~\ref{CURVACONDUCTIVIDAD} we plot the imaginary part of
$\tilde\sigma(\omega)$ as obtained from a field with $n_0=1$ and
$n_1=-1$, (from now on we shall only indicate the non-vanishing
$n_i$'s), in a system of Wilson fermions with $m=1/2$, $r=1$ and
$\mu=1$, that is  within the band energy-range and therefore with a
non-vanishing Fermi surface. We see that
for large frequencies the thermodynamic limit is reached in rather
small lattices. However, at the minimal reachable frequency ($2\pi/L_0$) the
conductivity is rapidly growing suggesting a singularity at zero
frequency. In fact, for a (classical) system of free particles of
density $n$ we expect that $\sigma(\omega)$ will behave as
\begin{equation}
\sigma^\mathrm{free,classical}\sim \ -{\mathrm i}\  
\frac{e^2 n}{m\, \omega}\, .
\end{equation}
Notice that if $\sigma(\omega)$ has a pole at $\omega=0$ with a purely
imaginary residue the same will hold true for $\tilde\sigma(\omega)$,
and both residues will be equal. Although the {\em Euclidean\/}
conductivity $\tilde\sigma(\omega)$ do not match the real-time one
(being imaginary, it cannot fulfill the Kramers-Kronig relations),
one can formally expect the residues to coincide in the passage from
$\omega$ to ${\mathrm i} \omega$.  This suggest to define the following
quantity which will be the basic object of our study:
\begin{equation}
Z^\mathrm{E}=
\frac{1}{i}\omega^\mathrm{min}\tilde\sigma(\omega^\mathrm{min})\,,\quad 
\omega^\mathrm{min}=\frac{2\pi}{L_0}\, .
\end{equation}
In the $L_i,L_0\to\infty$ limit, $Z^\mathrm{E}$ tend to the 
residue of the
pole. In order to measure this, we have considered the
smallest of possible external disturbances: $\{n_0=1,n_{L_0/2}=-1\}$.

\begin{figure}[t!]
\begin{center}
\leavevmode
\epsfig{file=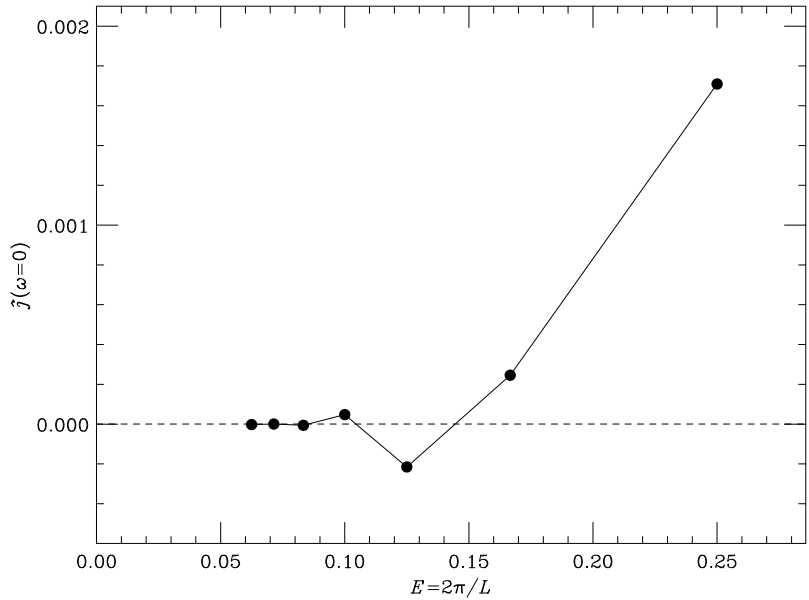,width=0.5\linewidth,angle=90}
\end{center}
\caption{\small The Fourier transform at zero frequency of the electrical
current as a function of the inverse lattice size.}
\label{NOLINEAL}
\end{figure}

Our result is shown in Fig. \ref{DEN-RES}, lower part.  We see
that the {\em Euclidean\/} residue follows quite closely Kohn's
result, which in fact can be considered as the infinite volume limit
for our calculation.  Moreover, the physical picture is rather
transparent: when the band is full, the system gets almost inert,
while when the band is empty, it can be excited by the external field
creating a hole in the Dirac sea. Since the smallest possible
excitation has frequency $2\pi/L_0$, to be compared with a gap $2m$,
it is reasonable that at $\mu=0$, the larger is the space-time
lattice, the smaller is the system response.  In fact, notice that in
Fig. \ref{DEN-RES} when $\mu$ is below the lower band limit, the
curves get horizontal: in this range of $\mu$ the system can be only
excited by crossing the gap between the Dirac sea and the conduction
band. And the gap is, of course, $\mu$ independent in a
non-interacting system.

We remark that our results have been found within the linear response
approximation.  We can control this approximation in several ways. One
is to study the Fourier transform of the current, for frequencies at
which the Fourier transform of the external-field vanishes.  In
Fig.~\ref{NOLINEAL} we show the zero-mode of the electrical current
for the electrical-field $\{n_0=1,n_1=-1\}$. We see that this
non-linear effect tends to zero with growing lattice-size, which is
quite reasonable since the minimum possible electric field is
$2\pi/L_1$. The non-linear corrections are oscillating, but modulated
by a rapidly decaying function.
Roughly speaking, for the largest lattice the non linear effects are
of the same order as the distance to the thermodynamic limit. A
further check can be done by comparing the residue obtained from the
data in Fig.~\ref{CURVACONDUCTIVIDAD} with the one in
Fig.~\ref{DEN-RES}: in the $L=14$ lattice, the differences are at the
$0.3\%$ level, while in the $L=6$ lattice the differences are at the
$1.6\%$ level.  Therefore, we believe that non-linear effects are
under control for the not too small fields that we can deal with.

\section{\protect\label{CONCLUSIONES}Conclusions}

We have presented a simple way of studying the electrical conductivity
of a system of Wilson fermions at finite density and zero-temperature
in a path-integral formalism.  In particular, we have
computed the residue of the zero frequency pole of the
conductivity, by numerically considering the linear response
to an external electric field, varying in Euclidean time. The results
have been contrasted with an analytical computation based on a method
proposed by Kohn, and an excellent agreement has been found.  As a
further cross-check, we have computed the density of states both
analytically and numerically in a finite lattice, obtaining a nice
thermodynamic limit convergence.  It should be emphasized that in
contrast with the analytical calculation which can only be done for a
non interacting system (or, at most, for simple external fields), 
the numerical calculations are easily generalizable to more
complex models, as fermions self-coupled with quartic interactions or
via a dynamic bosonic field.

An open, very interesting question is the possibility of extracting
the full real-time electrical conductivity function from its 
{\em Euclidean\/} counterpart. We have shown that the residue of the
zero-frequency pole can indeed be obtained. It would be also very
interesting to extend this approach to systems at finite-temperature. 

\section{Acknowledgements}

This work was triggered during a discussion with F. Guinea to whom we
are indebted for many suggestions, discussions and bibliographical
help. The numerical calculations have being carried-out in the RTNN
machines at the universities of Zaragoza and Complutense de Madrid.
This work has been partially supported by CICYT, contracts 
AEN97-1680,1693,1708.


\begin{thebibliography}{99}

\bibitem{LOMBARDO}F. Karsch, M-P. Lombardo (editors), 
{\em QCD at Finite Barion Density}, Nucl. Phys. {\bf
A642}(1998) Proc. Supp.  1-2.
\bibitem{FOURFERMIONS} F. Karsch, J.B. Kogut, H.W. Wyld,
Nucl. Phys. {\bf B280}[FS18](1987)289;
S.J. Hands, A. Koci\'c, J.B. Kogut, Nucl. Phys. {\bf B390}(1993)355;
S.J. Hands, S. Kim, J.B. Kogut, Nucl. Phys. {\bf B442}(1995)364;
I. Barbour, S. Hands, J.B. Kogut, M-P. Lombardo, S. Morrison, 
Nucl. Phys. {\bf B557}(1999)327.
\bibitem{SU2}
M.P. Lombardo, {\sl preprint} hep-lat/9907025;
S. Hands, J.B. Kogut, M.P. Lombardo, S. Morrison,
Nucl. Phys. {\bf B558}(1999)327.
\bibitem{KOHN} W. Kohn, Phys. Rev. {\bf 133}(1964)A171.
\bibitem{WILSON} K.G. Wilson, Phys. Rev. {\bf D14}(1974)2455.
\bibitem{WILSON2}K.G. Wilson. {\em Quarks and Strings on the
Lattice}, in New Phenomena in subnuclear Physics. Editor
A. Zichichi. Plenum Press, New York 1977.
\bibitem{POTQUIMH} P. Hasenfratz, F. Karsch, Phys. Lett. {\bf B125}(1983)308.
\bibitem{POTQUIMK} J.B. Kogut, H. Matsuoka, M. Stone, H.W. Wyld,
S. Shenker, J. Shigemitsum D.K. Sinclair, Nucl. Phys. {\bf B225}[FS9](1983)93.
\bibitem{BENDER} I. Bender, H.J. Rothe, I.O. Stomatescu, W. Wetzel,
Z. Phys. {\bf C58}(1993)333.
\bibitem{MONTVAYMUNSTER} I. Montvay, G. M\"unster, {\em Quantum Fields
on a Lattice}. Cambridge Univ. Press. 1994.
\bibitem{GAVAI} R.V. Gavai, Phys. Rev. {\bf D32}(1985)519.
\bibitem{ASCROFHTMERMIN} N.W. Ashcroft and N.D. Mermin, {\em Solid State
Physics}, Saunders College Publishing, 1976. 
\bibitem{EXTERNAL}
G. Martinelli, G. Parisi, R. Petronzio, F. Rapuano, Phys. Lett. 
{\bf 116B} (1982) 434;
C. Bernard, T. Draper, K. Olynyk, M. Rushton, Phys. Rev. Lett. 
{\bf 49} (1982) 1076;
J. Ambjorn, V.K. Mitryushkin, V. G. Bornyakov, A. M. Zadorozhnyi,
Phys. Lett. {\bf B225} (1989) 153;
M. L\"{u}scher, P. Weisz, R. Sommer, U. Wolff,
Nucl. Phys. {\bf B389} (1993) 247.

\end{thebibliography}
\end{document}